\documentclass[12pt]{article} 
\usepackage{amsmath}
\usepackage{amsfonts}
\usepackage{dsfont}
\usepackage{epsfig}
\usepackage{epic}
\usepackage{eepic}
\usepackage{axodraw}
\usepackage{subfigure}
\usepackage{rotating}

\def\arsh{\mbox{arsh}}
\def\arccot{\mbox{arccot}}

\def\be{\begin{equation}}
\def\ee{\end{equation}}
\newcommand{\bea}{\begin{eqnarray}}
\newcommand{\eea}{\end{eqnarray}}

\newlength{\diameter}
    
\newlength{\neglength}
         
\makeatletter
\newcommand{\redefinelabel}[1]{
  \def\@currentlabel{#1}}
\makeatother

\begin {document}
\begin{titlepage}
February 2003 \\
\begin{flushright}
HU Berlin-EP-03/07\\ 
\end{flushright}
\mbox{ }  \hfill hep-th/0302153
\vspace{5ex}
\Large
\begin {center}     
{\bf Conformal boundary and geodesics for $AdS_5\times S^5$ and the plane wave:
Their approach in the Penrose limit}
\end {center}
\large
\vspace{1ex}
\begin{center}
Harald Dorn and Christoph Sieg \footnote{dorn@physik.hu-berlin.de, csieg@physik.hu-berlin.de}
\end{center}
\begin{center}
Humboldt--Universit\"at zu Berlin, Institut f\"ur Physik\\
Invalidenstra\ss e 110, D-10115 Berlin\\[2mm]  
\end{center}
\vspace{4ex}
\rm
\begin{center}
{\bf Abstract}
\end{center} 
\normalsize 
Projecting on a suitable subset of coordinates, a picture is constructed
in which the conformal boundary of $AdS_5\times S^5$ and that of the plane 
wave resulting in the Penrose limit are located at the same line. In a second line of arguments all
$AdS_5\times S^5$ and plane wave geodesics are constructed in their integrated
form. Performing the Penrose limit, the approach of null geodesics reaching the conformal boundary of $AdS_5\times S^5$ to that of the plane wave is studied
in detail. At each point these null geodesics of $AdS_5\times S^5$ form a cone which degenerates in the limit.
\vfill
\end{titlepage} 
\section{Introduction}
The AdS/CFT correspondence relates ${\cal N} =4$ super Yang-Mills gauge theory
in Minkowski space to type IIB string theory in $AdS_5\times S^5$ with some 
RR background flux. But since the relevant string spectrum is available
only for large values of the t'Hooft coupling, explicit tests beyond
the supergravity approximation remain a very difficult task. Therefore,
a lot of activity has been induced by the observation of \cite{bmn},
that a certain sector of ${\cal N} =4$ super Yang-Mills, defined by a convenient restriction to operators with a large angular momentum in $S^5$ , can be 
related to string theory in a much simpler background, namely a plane wave
arising in some Penrose limit of the original background.
\cite{bfhp}.

One of the most puzzling questions in this framework is the issue of 
holography \cite{dgr,kp,lor,bn,s}. In the standard picture the ${\cal N} =4$ 
super Yang-Mills as the dual partner of the string theory resides on
the  conformal boundary of  $AdS_5\times S^5$ which is a
four-dimensional  Minkowski
space. The plane wave is generated by zooming into the neighbourhood
of a certain null geodesic in $AdS_5\times S^5$ followed by a constant 
rescaling of the metric. In this procedure the old boundary is put beyond
the new plane wave space. Nevertheless, string theoretic quantities referring
to the new plane wave are in well established correspondence to anomalous
dimensions and Green functions of a subsector of the old four-dimensional
field theory \cite{bmn,kpss,gmr,cfhmmps,sz}.

Obviously, a natural starting point for re-establishing some version
of the holographic picture centers around the analysis of the conformal
boundary of the plane wave. Via a suitable coordinate transformation
in \cite{bn} the plane wave was shown to be conformal to an Einstein
static universe $R\times S^9$. The conformal boundary was found, by inspection
of the singularities of the Weyl factor, to be an one-dimensional boundary
line. This analysis has been confirmed also using the more general
technique of terminal indecomposable pasts (futures) \cite{mr,hr}.\\

Given these results on the conformal boundary {\it after} the limit
has been 
performed, our aim in this paper is to shed some light on the behaviour
of the boundary in the limiting process itself. We follow the formulation
in \cite{bmn}, which presents the limit as a limit of the metric in given
coordinates. 

To this end the $AdS_5\times S^5$ metric in global coordinates 
\be
ds^2~=~R^2\big (-dt^2\cosh ^2\rho +d\rho ^2 +\sinh ^2\rho ~d\Omega ^2_3+
d\psi ^2\cos ^2\vartheta +d\vartheta ^2+\sin ^2\vartheta ~d\Omega _3^{
\prime ~2}\big )
\label{intro1}
\ee
via the replacements
\be
t~=~x^++\frac{x^-}{R^2}~,~~~\psi ~=~x^+-\frac{x^-}{R^2}~,~~~\rho~=~\frac{r}{R}~,~~~\vartheta ~=~\frac{y}{R}
\label{intro2}
\ee
is transformed to
\bea
ds^2&=&R^2(dx^+)^2(\cos ^2\frac{y}{R}-\cosh ^2\frac{r}{R})~-~2dx^+dx^-
(\cos ^2\frac{y}{R}+\cosh ^2\frac{r}{R})~\label{intro3} \\
&+&R^{-2}(dx^-)^2(\cos ^2\frac{y}{R}-\cosh ^2\frac{r}{R})+dr^2+dy^2
+R^2\sinh ^2\frac{r}{R}d\Omega _3^2~+R^2\sin ^2\frac{y}{R}d\Omega ^{\prime ~2}_3~.
\nonumber
\eea
Then in the limit $R\rightarrow\infty $ the metric becomes
\be
(ds^2)_{\mbox{\scriptsize pw}}~=~-4dx^+dx^-~-~\vec x^{~2}(dx^+)^2~+~(d\vec x)^2~.
\label{intro4}
\ee
Since we now have the same coordinates
for the plane wave and $AdS_5\times S^5$, we can address in section 2
the question about the relative location of the conformal boundary of both 
spaces. This issue will be discussed both in the coordinates just introduced
as well as in the coordinates used in \cite{bn} for the identification of the
conformal boundary of the plane wave.\footnote{For later reference we call 
them BMN and BN coordinates respectively.} To analyze the causal
structure of  the field theoretical holographic picture geodesics, and
in  particular null geodesics reaching the holographic screen out of
the bulk, play a central role. With this motivation
we present in sections 3 and 4 a complete classification  of geodesics
in $AdS_5\times S^5$ and the plane wave (\ref{intro4}) in their integrated
form. Because in the chosen framework the relevant Penrose limit
is realized as a limit of metrics depending on a parameter $R\rightarrow\infty $, it is obvious that locally $AdS_5\times S^5$ geodesics converge
to plane wave geodesics. But our knowledge from sections 3 and 4 will be
useful for global aspects. In particular it is used in section 5 for a 
discussion
of how differently null geodesics approach in both spaces the
respective conformal boundary and in which sense the $AdS_5\times S^5$
null geodesics reaching the $AdS_5\times S^5$ boundary approach in the
limit $R\rightarrow\infty $ null geodesics running to the plane
wave conformal boundary. Finally, section 6 is devoted to a summary
of the results and some conclusions.
\section{Common description of conformal boundaries of $AdS_5\times S^5$
and the BMN plane wave}
Since the angular coordinate $\psi $ in (\ref{intro1}) is constrained
by $-\pi \leq \psi\leq \pi$, the coordinates $x^+$ and $x^-$ are constrained
by
\be
R^2~x^+~-~\pi R^2~\leq ~x^-~\leq ~R^2~x^+~+~\pi R^2~.
\label{bound0}
\ee
This is a strip in the $(x^+,x^-)$-plane bounded by the two parallel 
straight lines with slope $R^2$ and crossing the $x^+$-axis at $-\pi$ and
$\pi$, respectively. For $R\rightarrow\infty$ this strip becomes
the coordinate range $-\infty <x^-<\infty ~,~~-\pi\leq x^+\leq\pi$.
Taking the limit for the metric, the identification of the two
boundaries of the strip is given up, and it makes sense to extend
to the whole $(x^+,x^-)$-plane. If one wants to avoid the restriction
to the strip already for finite $R$, one has to puncture $S^5$ at its poles
and to go then to the universal covering obtained by allowing $\psi $
to take any real value.

The sequence of coordinate transformations, done in \cite{bn} to analyze
the conformal boundary of the plane wave geometry (\ref{intro4}), can be
summarized as follows. Writing $(d\vec x)^2=x^2d\Omega _7^2$, the $\Omega _7$ coordinates remain untouched. 
Then in a first step one transforms 
\footnote{Note that our definitions for $x^{\pm}$ follow \cite{bmn} and thus
slightly differ from \cite{bn}.}
in the patch $x^+\in (-\frac{\pi}{2},\frac{\pi}{2})$ the coordinates $x^+,x^-,x$ to $\theta,\varphi,\zeta $
\be
\begin{aligned}
\cot\theta &=\frac{\big ((1-x^2)\tan x^+-4x^-\big )\cos x^+}{2x}~,\\
\tan\frac{\varphi \pm \zeta}{2}&=\frac{1}{2}(1+x^2)\tan x^+~+~2x^-~\pm ~\frac{x}{\sin\theta\cos x^+}~.
\end{aligned}
\label{bound1}
\ee
The  new coordinates are constrained by $0\leq\theta\leq\pi~,~0\leq \zeta\leq\pi ~,~\vert\varphi \pm
\zeta\vert\leq\pi $. The second step uses the periodicity properties
of the trigonometric functions to glue the other $x^+$-strips resulting
in the final coordinate range
\be
0\leq\theta\leq\pi~,~~~~0\leq \zeta\leq\pi ~,~~~-\infty <\varphi <\infty ~.
\label{bound2}
\ee
Then the plane wave metric, up to a conformal factor, turns out to be
that of the Einstein static universe $R\times S^9$. The analysis of singularities
of the conformal factor, determining the conformal boundary of the plane wave,
becomes most transparent after a change of parametrization of $S^9$. Let
denote $z_1,z_2,\vec z$ Cartesian coordinates in an embedding $R^{10}$,
then the parametrization by $\theta ,\zeta$ is related to that by $\alpha ,\beta$ via 
\footnote{We shift $\alpha $ to $\alpha -\frac{\pi}{2}$ and $\beta $ to $\beta -\pi$ relative to \cite{bn}.}
\bea
z_1&=&\cos\zeta ~=~\sin\alpha ~\cos\beta ~,\nonumber\\
z_2&=&\cos\theta ~\sin\zeta ~=~\sin\alpha ~\sin\beta ~,\nonumber\\
\vert\vec z\vert &=& \sin\theta~\sin\zeta ~=~\cos\alpha~.
\label{bound3}
\eea 
The range for $\alpha ,\beta $ is
\be
0\leq\alpha\leq\frac{\pi}{2}~,~~~0\leq\beta\leq 2\pi~.
\label{bound4}
\ee  
Now the plane wave metric in these BN coordinates takes the form \cite{bn}
\be
(ds^2)_{\mbox{\scriptsize pw}}~=~\frac{1}{\vert e^{i\varphi}+\sin\alpha ~e^{i\beta}\vert ^2}\left (-d\varphi ^2+d\alpha ^2+\sin ^2\alpha ~d\beta ^2+\cos ^2\alpha ~d\Omega _7^2\right )~.
\label{bound5}
\ee
The conformal factor is singular iff $\alpha =\frac{\pi}{2}$ and $\varphi =\beta +(2k+1)\pi ~,k\in ${\bf Z}. Since at $\alpha =\frac{\pi}{2}$ \\the $S^7$
part due the $\cos ^2\alpha $ factor in front of $d\Omega _7^2$ \ shrinks
to a point, the conformal boundary
\footnote{Here and for the $AdS_5\times S^5$ case below, while speaking
about the conformal boundary, we omit the two isolated points for
timelike infinity.}
 of the plane wave is one-dimensional, see also fig.~1.\\
\begin {figure}
\begin{center}
\setlength{\unitlength}{0.240900pt}
\begin{picture}(1500,900)(0,0)
\footnotesize
\put(762,649){\makebox(0,0)[l]{$\scriptstyle\alpha=0$}}
\put(958,649){\makebox(0,0)[l]{$\scriptstyle\alpha=\frac{\pi}{2}$}}
\put(803,457){\makebox(0,0)[l]{$\scriptstyle\delta$}}
\put(749,763){\makebox(0,0)[l]{$\varphi$}}
\thinlines \path(749,472)(939,458)
\put(939,458){\line(0,1){45}}
\thinlines \path(749,472)(848,427)
\put(945,294){\line(0,1){355}}
\put(554,294){\line(0,1){355}}
\thinlines \path(609,535)(890,408)\put(890,408){\vector(2,-1){0}}

\thinlines \path(506,435)(993,508)\put(993,508){\vector(4,1){0}}

\put(749,280){\vector(0,1){458}}
\thinlines \path(651,338)(631,342)
\thinlines \path(631,342)(612,345)
\thinlines \path(612,345)(595,347)
\thinlines \path(595,347)(581,349)
\thinlines \path(581,349)(570,350)
\thinlines \path(570,350)(561,351)
\thinlines \path(561,351)(556,352)
\thinlines \path(556,352)(554,353)
\thinlines \path(554,353)(554,354)
\thinlines \path(554,354)(558,354)
\thinlines \path(558,354)(565,355)
\thinlines \path(565,355)(575,357)
\thinlines \path(575,357)(588,358)
\thinlines \path(588,358)(603,360)
\thinlines \path(603,360)(621,363)
\thinlines \path(621,363)(641,366)
\thinlines \path(641,366)(662,370)
\thinlines \path(662,370)(685,375)
\thinlines \path(685,375)(709,380)
\thinlines \path(709,380)(734,386)
\thinlines \path(734,386)(759,393)
\thinlines \path(759,393)(783,401)
\thinlines \path(783,401)(807,410)
\thinlines \path(807,410)(831,419)
\thinlines \path(831,419)(853,430)
\thinlines \path(853,430)(873,441)
\thinlines \path(873,441)(891,452)
\thinlines \path(891,452)(907,464)
\thinlines \path(907,464)(921,477)
\thinlines \path(921,477)(931,490)
\thinlines \path(931,490)(939,503)
\thinlines \path(939,503)(944,517)
\thinlines \path(944,517)(946,530)
\thinlines \path(946,530)(944,544)
\thinlines \path(944,544)(939,558)
\thinlines \path(939,558)(932,571)
\thinlines \path(932,571)(921,584)
\thinlines \path(921,584)(908,597)
\thinlines \path(908,597)(892,609)
\thinlines \path(892,609)(874,620)
\thinlines \path(874,620)(853,631)
\thinlines \path(853,631)(831,642)
\thinlines \path(831,642)(808,651)
\thinlines \path(808,651)(784,660)
\thinlines \path(784,660)(759,668)
\thinlines \path(759,668)(735,675)
\thinlines \path(735,675)(710,681)
\thinlines \path(710,681)(686,687)
\thinlines \path(686,687)(663,691)
\thinlines \path(663,691)(652,693)
\thinlines \path(651,693)(631,690)
\thinlines \path(631,690)(612,686)
\thinlines \path(612,686)(595,681)
\thinlines \path(595,681)(581,675)
\thinlines \path(581,675)(570,670)
\thinlines \path(570,670)(561,663)
\thinlines \path(561,663)(556,657)
\thinlines \path(556,657)(554,651)
\thinlines \path(554,651)(554,644)
\thinlines \path(554,644)(558,638)
\thinlines \path(558,638)(565,632)
\thinlines \path(565,632)(575,626)
\thinlines \path(575,626)(588,620)
\thinlines \path(588,620)(603,615)
\thinlines \path(603,615)(621,610)
\thinlines \path(621,610)(641,606)
\thinlines \path(641,606)(662,603)
\thinlines \path(662,603)(685,601)
\thinlines \path(685,601)(709,599)
\thinlines \path(709,599)(734,598)
\thinlines \path(734,598)(759,598)
\thinlines \path(759,598)(783,599)
\thinlines \path(783,599)(807,600)
\thinlines \path(807,600)(831,602)
\thinlines \path(831,602)(853,605)
\thinlines \path(853,605)(873,609)
\thinlines \path(873,609)(891,614)
\thinlines \path(891,614)(907,619)
\thinlines \path(907,619)(921,624)
\thinlines \path(921,624)(931,630)
\thinlines \path(931,630)(939,636)
\thinlines \path(939,636)(944,642)
\thinlines \path(944,642)(946,649)
\thinlines \path(946,649)(944,655)
\thinlines \path(944,655)(939,662)
\thinlines \path(939,662)(932,668)
\thinlines \path(932,668)(921,674)
\thinlines \path(921,674)(908,679)
\thinlines \path(908,679)(892,684)
\thinlines \path(892,684)(874,689)
\thinlines \path(874,689)(853,693)
\thinlines \path(853,693)(831,696)
\thinlines \path(831,696)(808,698)
\thinlines \path(808,698)(784,699)
\thinlines \path(784,699)(759,700)
\thinlines \path(759,700)(735,700)
\thinlines \path(735,700)(710,699)
\thinlines \path(710,699)(686,698)
\thinlines \path(686,698)(663,695)
\thinlines \path(663,695)(641,692)
\thinlines \path(641,692)(622,688)
\thinlines \path(622,688)(604,683)
\thinlines \path(604,683)(588,678)
\thinlines \path(588,678)(576,673)
\thinlines \path(576,673)(566,667)
\thinlines \path(566,667)(558,661)
\thinlines \path(558,661)(554,654)
\thinlines \path(554,654)(554,648)
\thinlines \path(554,648)(556,641)
\thinlines \path(556,641)(561,635)
\thinlines \path(561,635)(570,629)
\thinlines \path(570,629)(581,623)
\thinlines \path(581,623)(595,618)
\thinlines \path(595,618)(611,613)
\thinlines \path(611,613)(630,608)
\thinlines \path(630,608)(651,605)
\thinlines \path(651,605)(673,602)
\thinlines \path(673,602)(696,600)
\thinlines \path(696,600)(721,598)
\thinlines \path(721,598)(746,598)
\thinlines \path(746,598)(770,598)
\thinlines \path(770,598)(795,599)
\thinlines \path(795,599)(819,601)
\thinlines \path(819,601)(841,604)
\thinlines \path(841,604)(862,607)
\thinlines \path(862,607)(882,611)
\thinlines \path(882,611)(899,616)
\thinlines \path(899,616)(914,621)
\thinlines \path(914,621)(926,627)
\thinlines \path(926,627)(935,633)
\thinlines \path(935,633)(942,639)
\thinlines \path(942,639)(945,646)
\thinlines \path(945,646)(945,652)
\thinlines \path(945,652)(942,658)
\thinlines \path(942,658)(936,665)
\thinlines \path(936,665)(927,671)
\thinlines \path(927,671)(915,676)
\thinlines \path(915,676)(900,682)
\thinlines \path(900,682)(883,686)
\thinlines \path(883,686)(864,691)
\thinlines \path(864,691)(843,694)
\thinlines \path(843,694)(821,697)
\thinlines \path(821,697)(797,699)
\thinlines \path(797,699)(772,700)
\thinlines \path(772,700)(748,700)
\thinlines \path(748,700)(723,700)
\thinlines \path(723,700)(698,699)
\thinlines \path(698,699)(675,696)
\thinlines \path(675,696)(653,694)
\thinlines \path(651,516)(631,512)
\thinlines \path(631,512)(612,508)
\thinlines \path(612,508)(595,503)
\thinlines \path(595,503)(581,498)
\thinlines \path(581,498)(570,492)
\thinlines \path(570,492)(561,486)
\thinlines \path(561,486)(556,480)
\thinlines \path(556,480)(554,473)
\thinlines \path(554,473)(554,467)
\thinlines \path(554,467)(558,460)
\thinlines \path(558,460)(565,454)
\thinlines \path(565,454)(575,448)
\thinlines \path(575,448)(588,442)
\thinlines \path(588,442)(603,437)
\thinlines \path(603,437)(621,433)
\thinlines \path(621,433)(641,429)
\thinlines \path(641,429)(662,426)
\thinlines \path(662,426)(685,423)
\thinlines \path(685,423)(709,421)
\thinlines \path(709,421)(734,420)
\thinlines \path(734,420)(759,420)
\thinlines \path(759,420)(783,421)
\thinlines \path(783,421)(807,423)
\thinlines \path(807,423)(831,425)
\thinlines \path(831,425)(853,428)
\thinlines \path(853,428)(873,432)
\thinlines \path(873,432)(891,436)
\thinlines \path(891,436)(907,441)
\thinlines \path(907,441)(921,447)
\thinlines \path(921,447)(931,452)
\thinlines \path(931,452)(939,459)
\thinlines \path(939,459)(944,465)
\thinlines \path(944,465)(946,471)
\thinlines \path(946,471)(944,478)
\thinlines \path(944,478)(939,484)
\thinlines \path(939,484)(932,490)
\thinlines \path(932,490)(921,496)
\thinlines \path(921,496)(908,502)
\thinlines \path(908,502)(892,507)
\thinlines \path(892,507)(874,511)
\thinlines \path(874,511)(853,515)
\thinlines \path(853,515)(831,518)
\thinlines \path(831,518)(808,520)
\thinlines \path(808,520)(784,522)
\thinlines \path(784,522)(759,523)
\thinlines \path(759,523)(735,523)
\thinlines \path(735,523)(710,522)
\thinlines \path(710,522)(686,520)
\thinlines \path(686,520)(663,517)
\thinlines \path(663,517)(641,514)
\thinlines \path(641,514)(622,510)
\thinlines \path(622,510)(604,506)
\thinlines \path(604,506)(588,501)
\thinlines \path(588,501)(576,495)
\thinlines \path(576,495)(566,489)
\thinlines \path(566,489)(558,483)
\thinlines \path(558,483)(554,477)
\thinlines \path(554,477)(554,470)
\thinlines \path(554,470)(556,464)
\thinlines \path(556,464)(561,457)
\thinlines \path(561,457)(570,451)
\thinlines \path(570,451)(581,445)
\thinlines \path(581,445)(595,440)
\thinlines \path(595,440)(611,435)
\thinlines \path(611,435)(630,431)
\thinlines \path(630,431)(651,427)
\thinlines \path(651,427)(673,424)
\thinlines \path(673,424)(696,422)
\thinlines \path(696,422)(721,421)
\thinlines \path(721,421)(746,420)
\thinlines \path(746,420)(770,421)
\thinlines \path(770,421)(795,422)
\thinlines \path(795,422)(819,424)
\thinlines \path(819,424)(841,426)
\thinlines \path(841,426)(862,430)
\thinlines \path(862,430)(882,434)
\thinlines \path(882,434)(899,438)
\thinlines \path(899,438)(914,444)
\thinlines \path(914,444)(926,449)
\thinlines \path(926,449)(935,455)
\thinlines \path(935,455)(942,462)
\thinlines \path(942,462)(945,468)
\thinlines \path(945,468)(945,474)
\thinlines \path(945,474)(942,481)
\thinlines \path(942,481)(936,487)
\thinlines \path(936,487)(927,493)
\thinlines \path(927,493)(915,499)
\thinlines \path(915,499)(900,504)
\thinlines \path(900,504)(883,509)
\thinlines \path(883,509)(864,513)
\thinlines \path(864,513)(843,517)
\thinlines \path(843,517)(821,519)
\thinlines \path(821,519)(797,521)
\thinlines \path(797,521)(772,522)
\thinlines \path(772,522)(748,523)
\thinlines \path(748,523)(723,522)
\thinlines \path(723,522)(698,521)
\thinlines \path(698,521)(675,519)
\thinlines \path(675,519)(653,516)
\thinlines \path(651,338)(631,335)
\thinlines \path(631,335)(612,331)
\thinlines \path(612,331)(595,326)
\thinlines \path(595,326)(581,320)
\thinlines \path(581,320)(570,315)
\thinlines \path(570,315)(561,308)
\thinlines \path(561,308)(556,302)
\thinlines \path(556,302)(554,296)
\thinlines \path(554,296)(554,289)
\thinlines \path(554,289)(558,283)
\thinlines \path(558,283)(565,276)
\thinlines \path(565,276)(575,271)
\thinlines \path(575,271)(588,265)
\thinlines \path(588,265)(603,260)
\thinlines \path(603,260)(621,255)
\thinlines \path(621,255)(641,251)
\thinlines \path(641,251)(662,248)
\thinlines \path(662,248)(685,246)
\thinlines \path(685,246)(709,244)
\thinlines \path(709,244)(734,243)
\thinlines \path(734,243)(759,243)
\thinlines \path(759,243)(783,243)
\thinlines \path(783,243)(807,245)
\thinlines \path(807,245)(831,247)
\thinlines \path(831,247)(853,250)
\thinlines \path(853,250)(873,254)
\thinlines \path(873,254)(891,259)
\thinlines \path(891,259)(907,264)
\thinlines \path(907,264)(921,269)
\thinlines \path(921,269)(931,275)
\thinlines \path(931,275)(939,281)
\thinlines \path(939,281)(944,287)
\thinlines \path(944,287)(946,294)
\thinlines \path(946,294)(944,300)
\thinlines \path(944,300)(939,307)
\thinlines \path(939,307)(932,313)
\thinlines \path(932,313)(921,319)
\thinlines \path(921,319)(908,324)
\thinlines \path(908,324)(892,329)
\thinlines \path(892,329)(874,334)
\thinlines \path(874,334)(853,337)
\thinlines \path(853,337)(831,341)
\thinlines \path(831,341)(808,343)
\thinlines \path(808,343)(784,344)
\thinlines \path(784,344)(759,345)
\thinlines \path(759,345)(735,345)
\thinlines \path(735,345)(710,344)
\thinlines \path(710,344)(686,342)
\thinlines \path(686,342)(663,340)
\thinlines \path(663,340)(641,337)
\thinlines \path(641,337)(622,333)
\thinlines \path(622,333)(604,328)
\thinlines \path(604,328)(588,323)
\thinlines \path(588,323)(576,318)
\thinlines \path(576,318)(566,312)
\thinlines \path(566,312)(558,305)
\thinlines \path(558,305)(554,299)
\thinlines \path(554,299)(554,293)
\thinlines \path(554,293)(556,286)
\thinlines \path(556,286)(561,280)
\thinlines \path(561,280)(570,274)
\thinlines \path(570,274)(581,268)
\thinlines \path(581,268)(595,262)
\thinlines \path(595,262)(611,258)
\thinlines \path(611,258)(630,253)
\thinlines \path(630,253)(651,250)
\thinlines \path(651,250)(673,247)
\thinlines \path(673,247)(696,245)
\thinlines \path(696,245)(721,243)
\thinlines \path(721,243)(746,243)
\thinlines \path(746,243)(770,243)
\thinlines \path(770,243)(795,244)
\thinlines \path(795,244)(819,246)
\thinlines \path(819,246)(841,249)
\thinlines \path(841,249)(862,252)
\thinlines \path(862,252)(882,256)
\thinlines \path(882,256)(899,261)
\thinlines \path(899,261)(914,266)
\thinlines \path(914,266)(926,272)
\thinlines \path(926,272)(935,278)
\thinlines \path(935,278)(942,284)
\thinlines \path(942,284)(945,290)
\thinlines \path(945,290)(945,297)
\thinlines \path(945,297)(942,303)
\thinlines \path(942,303)(936,310)
\thinlines \path(936,310)(927,316)
\thinlines \path(927,316)(915,321)
\thinlines \path(915,321)(900,327)
\thinlines \path(900,327)(883,331)
\thinlines \path(883,331)(864,336)
\thinlines \path(864,336)(843,339)
\thinlines \path(843,339)(821,342)
\thinlines \path(821,342)(797,344)
\thinlines \path(797,344)(772,345)
\thinlines \path(772,345)(748,345)
\thinlines \path(748,345)(723,345)
\thinlines \path(723,345)(698,343)
\thinlines \path(698,343)(675,341)
\thinlines \path(675,341)(653,339)
\thinlines \path(812,443)(825,446)
\thinlines \path(825,446)(837,448)
\thinlines \path(837,448)(848,451)
\thinlines \path(848,451)(857,455)
\thinlines \path(857,455)(864,458)
\thinlines \path(864,458)(869,462)
\end{picture}

\vspace{-1.5cm}
\end{center}
\noindent {\bf Fig.~1}\ \ {\it Part of the boundary of $AdS_5\times
  S^5$ and of  the plane wave in BN coordinates ($\delta =\beta+\pi$)}
\end {figure}

To avoid confusion in comparing fig.~1 with similar looking pictures for
$AdS_5\times S^5$, where the half of some Einstein static universe is
depicted, it is appropriate to stress that fig.~1 represents the
whole Einstein static universe $R\times S^9$ although the radius variable
of the cylinder runs from zero to $\frac{\pi}{2}$ only. This range for $\alpha$
is due to its special role in the parametrization of $S^9$ in (\ref{bound3}).\\

The coordinate transformations just discussed for the identification
of the conformal boundary of the plane wave of course can also be applied 
to the $AdS_5\times S^5$ metric. A priori these new coordinates are not 
a favourite choice to give any special insight into the $AdS_5\times S^5$ geometry. In particular they are not well suited to find the conformal boundary. 

But we can turn the argument around. Since we know already the conformal
boundary of $AdS_5\times S^5$, we can look where this boundary is situated
in the new coordinates and hope to find some illuminating picture for its degeneration in the $R\rightarrow\infty$ limit which produces the plane wave metric.

Starting with global coordinates (\ref{intro1}), the conformal 
$AdS_5\times S^5$ boundary is at $\rho\rightarrow\infty $ with all other
coordinates kept fixed at arbitrary finite values. Translating this into the
coordinates of (\ref{intro3}) it is at $r\rightarrow\infty$ and $x^+,x^-,y,\Omega
_3,\Omega ^{\prime}_3$ fixed at arbitrary finite values. Before applying    
(\ref{bound1}) we define $x$ for the $AdS_5\times S^5$ case
by
\be
r~=~x\cos\omega ~,~~~y~=~x\sin\omega ~,
\label{bound6}
\ee
i.e. $x^2=r^2+y^2$.
In the following coordinate transformation according to (\ref{bound1}) $\omega ,\Omega _3,\Omega ^{\prime}_3$ remain untouched. The conformal $AdS_5\times S^5$ boundary is now at $x\rightarrow\infty ,~\omega\rightarrow 0$ and $x^+,x^-,
\Omega _3,\Omega ^{\prime}_3$ fixed at arbitrary finite values.
Then from the first equation (\ref{bound1}) one finds (as above we again start with $\vert x^+\vert\leq\frac{\pi}{2}$ and glue the other $x^+$ patches afterwards)
\be
\lim _{x\rightarrow\infty ,~x^+>0}\theta ~=~\pi ~,~~~\lim _{x\rightarrow\infty ,~x^+<0}\theta ~=~0~,~~~\lim _{x\rightarrow\infty ,~x^+=0}\theta ~=~\frac{\pi}{2}~.
\label{bound7}
\ee 
Furthermore, by coupling $x^+\rightarrow 0$ in a suitable way with $x\rightarrow\infty $ one can reach any $\theta\in (0,\pi)$
\be
\lim _{x\rightarrow\infty ,~x^+=c/x}\theta ~=~\arctan (-2c^{-1})~.
\label{bound8}
\ee
The second equation of (\ref{bound1}) then yields
\be
\begin{aligned}
&x\rightarrow\infty ~,~\left\{\begin{array}{c}x^+<0 ~~(\mbox{i.e.}~\theta \rightarrow 0)\\
x^+>0~~(\mbox{i.e.}~\theta \rightarrow \pi)\end{array}\right\}\\
&\qquad\qquad\qquad\Longrightarrow ~\tan \left(\frac{\varphi+\zeta}{2}\right )\rightarrow\left\{\begin{array}{c}\mbox{finite}\\\infty\end{array}\right\}
~,~~\tan \left(\frac{\varphi-\zeta}{2}\right )\rightarrow\left\{\begin{array}{c}\infty\\\mbox{finite}\end{array}\right\}~.
\end{aligned}
\label{bound9}
\ee  
In addition one gets for $x\rightarrow\infty $ coupled as in (\ref{bound8})
with $x^+\rightarrow 0$
\be
x\rightarrow\infty ~,~~~0<\theta <\pi ~~~~\Longrightarrow ~~~\tan\left (\frac{\varphi\pm\zeta}{2}\right )~\rightarrow\pm\infty ~.
\label{bound10}
\ee
Putting together (\ref{bound7})-(\ref{bound10}), we see that in the projection
onto the three coordinates $\varphi ,\theta ,\zeta$ the conformal boundary
of the ($\vert x^+\vert <\frac{\pi}{2}$)-patch of $AdS_5\times S^5$
is mapped to the one-dimensional line starting at $(\varphi ,\zeta ,\theta)
=(-\pi ,0,0)$, running first with $\theta = 0$ and $\zeta -\varphi =\pi$
to $(\varphi ,\zeta ,\theta)=(0 ,\pi,0)$, then with $\varphi =0$ and $\zeta =\pi$
to $(\varphi ,\zeta ,\theta)=(0 ,\pi,\pi)$ and finally with $\theta =\pi $ and
$\varphi +\zeta =\pi$ to $(\varphi ,\zeta ,\theta)=(\pi ,0,\pi)$, see also
fig.~2.
\begin{figure}
\begin{center}
\begin{picture}(150,230)(0,0)
\SetOffset(60,120)
\LongArrow(0,0)(-35,-35)\Text(-40,-40)[]{$\zeta$}
\LongArrow(0,0)(60,0)\Text(70,0)[]{$\theta$}
\LongArrow(0,0)(0,90)\Text(0,100)[]{$\varphi$}
\DashLine(40,-90)(40,90){4}\Text(45,-5)[]{$\scriptsize\pi$}
\DashLine(0,-90)(0,90){4}\Text(-7.5,40)[]{$\scriptsize\pi$}
\DashLine(-20,-110)(-20,70){4}\Text(-25,-15)[]{$\scriptsize\pi$}
\DashLine(20,-110)(20,70){4}
\SetWidth{1}
\DashLine(-20,60)(0,40){2}
\DashLine(0,40)(40,40){2}
\Line(40,40)(20,-20)
\Line(20,-20)(-20,-20)
\Line(-20,-20)(0,-40)
\DashLine(0,-40)(40,-40){2}
\DashLine(40,-40)(20,-100){2}
\end{picture}
\end{center}
\noindent {\bf Fig.~2}\ \ {\it Part of the boundary of $AdS_5\times
  S^5$ and of the plane wave in $(\varphi,\zeta,\theta)$ coordinates}
\end{figure}

Translating this via (\ref{bound3}) into the coordinates $\varphi ,~\alpha ,
~\beta $ we find the line\footnote{Note that the piece  from $(\varphi ,\zeta ,\theta)=(0 ,\pi,0)$ to $(\varphi ,\zeta ,\theta)=(0 ,\pi,\pi)$ with $\varphi =0$ and $\zeta =\pi$ is mapped to one point $(\varphi ,\alpha ,\beta )=(0,\frac{\pi}{2},\pi )$.} 
$\alpha ~=~\frac{\pi}{2}$, $\beta ~=~\pi ~+~\varphi$, $-\pi <\varphi <+~\pi$.
After gluing the other $x^+$-patches we can conclude:

The projection onto the coordinates $(\varphi ,\alpha ,\beta )$ of the conformal boundary of $AdS_5\times S^5$ coincides with that of a part of the conformal boundary of the plane wave (\ref{intro4}). That in this projection only a part of the plane wave boundary line appears as the $AdS_5\times S^5$ boundary is due to the restriction to the $AdS_5\times S^5$-strip (\ref{bound0}). Note that
this restriction can be circumvented as discussed at the beginning of this section.

Taking into account the other seven coordinates, the $AdS_5\times S^5$ boundary
is of course not one-dimensional. But by using the same coordinates
both for $AdS_5\times S^5$ $and$ the plane wave, we now have visualized
the degeneration of the conformal boundary in the process of approaching the
plane wave limit. In the projection to three of the BN coordinates ($\varphi ,~\alpha ,~\beta $) the boundary stays throughout this process at the same location. The extension in the remaining coordinates degenerates to a point in the limit.\\

The picture is more involved if one compares the two boundaries in the
BMN coordinates $(x^+,x^-,x)$. As noted in \cite{bn}, due to the singularity
of the coordinate transformation on the boundary line in $(\varphi ,\alpha ,\beta)$, the limits in $(x^+,x^-,x)$ which map to this boundary line are not unique. Besides
\be
\mbox{Limit (i):}~~~x~\rightarrow ~\infty ~,~~~x^+,~x^-~~\mbox{finite}~,
\label{bound11}
\ee
just discussed above, the second limit is
\be
\mbox{Limit (ii):}~~~x^-~\rightarrow ~\pm ~\infty ~,~~~x^+,~x~~\mbox{finite}~,
\label{bound12}
\ee
as trivially seen from (\ref{bound1}).

Now the situation looks a little bit cumbersome. The conformal boundary  of $AdS_5\times S^5$  is realized via limit (i), that of the plane wave via both
limits (i) $and$ (ii), although the the plane wave itself is a limit of
 $AdS_5\times S^5$.

To get a better understanding of this situation, we are now asking what
happens in the $R\rightarrow\infty $ limit with geodesics, in particular null geodesics, which reach the conformal $AdS_5\times S^5$ boundary. After an explicit construction of all the  geodesics both for $AdS_5\times S^5$ and the plane wave (\ref{intro4}) in the next two sections, we come back to this question in section 5.
\section{Geodesics in $AdS_5\times S^5$}
We start with the $AdS_5\times S^5$ metric in global coordinates
\be
ds^2~=~R^2(-dt^2\cosh ^2\rho ~+~d\rho ^2 ~+~\sinh ^2\rho ~d\Omega ^2_3~+~d\Omega ^2_5)~.\label{geo1}
\ee
The geodesic equations for the $AdS_5$ and $S^5$ coordinates decouple. Geodesics on $S^5$ are great circles. Whether the geodesic in the total manifold 
$AdS_5\times S^5$ moves
in $S^5$ or stays at a fixed $S^5$ position has consequences for the overall causal property (space-like, time-like, null) only. There is no effect on the $AdS_5$
coordinates. Therefore we can concentrate on the $AdS_5$ part. Then
the geodesic equations for $t,\rho $ appearing in (\ref{geo1}) explicitly,
as well as  for $\phi ^i,~i=1,2,3$ being orthogonal coordinates on the 3-dim
unit sphere $\Omega _3$, 
are 
\bea
\ddot t~+~2\dot{\rho}~\dot t\tanh\rho ~=~0~,\label{geo2}\\
\ddot{\rho}~+~(\dot t\dot t ~-~\dot{\phi} _i \dot{\phi} _i h_i)
\sinh\rho\cosh\rho ~=~0~,\label{geo3}
\eea
\be
\ddot{\phi}^i~+~2\dot{\phi}^i\dot{\rho}\coth\rho~+~\dot{\phi}^k\dot{\phi}^j
\gamma _{kj}^i~=~0~.
\label{geo4}
\ee
The dot indicates differentiation with respect to an affine parameter $\tau $.
$h_i$ denotes the diagonal entries of the $\Omega _3$ metric and the  $\gamma _{kj}^i$ are the related Levi-Civita connection coefficients.

Since the radius of the $S^3$ contribution in (\ref{geo1}) depends on $\rho $,
the separation of the movement in $S^3$ is not as trivial as the separation of the $S^5$ movement. However, the $\rho$-dependence in eq.~(\ref{geo4}) 
appears as a factor of $\dot{\phi}_i$ only. This means that this equation
describes within $\Omega _3$ a geodesic, but our parameter $\tau $, defined to be
an affine one with respect to the geodesic as a whole, is not affine
for the $\Omega _3$ part treated separately on its own. Defining a new
parameter $\sigma ~=~f(\tau )$ as a solution of the differential equation
\be
\ddot f~+~2\dot{\rho}\dot f\coth\rho ~=~0~,
\label{geo5}
\ee
eq.~(\ref{geo4}) becomes equivalent to (the prime stands for $\frac{d}{d\sigma }$)
\be
{\phi}''^i~+~{\phi}'^k{\phi}'^j \gamma _{kj}^i~=~0~.
\label{geo6}
\ee
Now we are sure that the length of the tangential vector $\phi '^i$
in terms of the 3-dim unit sphere metric is constant, i.e.
$\phi '^i\phi '^i h_i~=~c_3\geq 0$. Transforming this back to $\dot{\phi}^i$
we get
\be 
\dot{\phi}^i\dot{\phi}^ih_i~=~c_3~\dot f^2~,~~~c_3\geq 0~.
\label{geo7}
\ee 
The freedom to multiply an affine parameter by a constant can be used
to choose $c_3=1$ or $c_3=0$ (for $\phi ^i=\mbox{const}$). However, in eq.~(\ref{geo7})
the option $c_3=0$ is no longer necessary, since constant $\phi ^i$
can be realized via $\dot f=0$. Therefore, from now on we put $c_3=1$ for
all cases. Inserting then (\ref{geo7}) into (\ref{geo3}) we arrive at
\be
\ddot{\rho}~+~(\dot t\dot t ~-~\dot f\dot f)\sinh\rho\cosh\rho ~=~0~,
\label{geo8}
\ee  
and see that the movement in $\rho$ and $t$ is influenced by a possible
movement within the unit 3-sphere coordinates of  $AdS_5$ only via one function $f(\tau )$
which on its own couples back to $\rho$ via (\ref{geo5}). 

Summarizing the discussion so far, the coordinates in $\Omega _3$ either
remain constant ($\dot f=0$) or describe a movement on a great circle ($
\dot f\neq 0$).
In addition, the system of five coupled equations for the coordinates 
$\rho ,t,\phi ^1,\phi ^2,\phi ^3$ is
reduced to the three coupled equations (\ref{geo2}),(\ref{geo8}) and (\ref{geo5}) for $\rho ,t,f$.

Straightforward integration of (\ref{geo2}) and (\ref{geo5}) yields  
\be
\dot t~=~\frac{b}{\cosh ^2\rho}~,~~~~~~\dot f~=~\frac{\tilde b}{\sinh ^2\rho}~,
~~~~~~b,~\tilde b ~~  \mbox{constant}~.
\label{geo9}
\ee
With (\ref{geo8}) this gives an equation for $\rho$ alone
\be
\ddot{\rho}~+~\frac{b^2}{\cosh ^3\rho}\sinh\rho ~-~\frac{\tilde b^2}{\sinh ^3\rho}\cosh\rho ~=~0~.
\label{geo10}
\ee
Instead of solving this equation directly, we found it more convenient to make
a small detour. Since our parameter $\tau $ is an affine one, the scalar
product of the tangential vector with itself in the sense of the $AdS$ metric
is a constant along the geodesic, we call it $c_5$. Then using 
(\ref{geo7}),(\ref{geo9}) we get
\be
\dot{\rho}^2~=~\frac{c_5}{R^2}~-~\frac{\tilde b^2}{\sinh ^2\rho}~+\frac{b^2}{\cosh ^2\rho}~.
\label{geo11}
\ee
Are all solutions of (\ref{geo11}) also solutions of (\ref{geo10})?
At first the constancy of the scalar product of the tangential vector with itself is of course a much weaker condition than the geodesic equations. But in
writing down (\ref{geo11}) we already have implied the geodesic equations
for all coordinates, except for $\rho$. Under these circumstances, at least
as long as $\dot\rho\neq 0$, the constant scalar product condition
is equivalent to the geodesic equation for the last coordinate $\rho$.\\

Since $\dot{\rho}^2$ is a non-negative quantity, from (\ref{geo11}) and
$$\frac{c_5}{R^2}~-~\frac{\tilde b^2}{\sinh ^2\rho}~+\frac{b^2}{\cosh ^2\rho}~\leq ~\frac{c_5}{R^2}~+~b^2~,~~~\forall\rho ~,$$
as a byproduct, we find a constraint on $c_5$ and the integration constant $b$
\be
\frac{c_5}{R^2}~+~b^2~\geq ~0~.\label{geo12}
\ee

For further analyzing the consequences of the positiveness of both sides of eq.~(\ref{geo11}) we introduce the abbreviations
\be
A~=~\frac{c_5}{R^2}~,~~~~B~=~b^2~+~\frac{c_5}{R^2}-\tilde b^2~,~~~~C~=~-\tilde b^2~.
\label{geo13}
\ee
Then first of all, by these definitions and the inequality (\ref{geo12}) the constants 
$A,B,C$ are universally constrained by
\be
C~\leq ~0~,~~~B~\geq ~A~+~C~,~~~B~\geq ~C~.
\label{geo13a}
\ee
In addition, checking whether there are real $\rho $-values for which
the r.h.s. of (\ref{geo11}) is non-negative, it turns out that only
four classes of ranges\footnote{The special case $A=B=C=0$ corresponds to a point, not to a curve.}
of the constants $A,B,C$  are allowed. Integrating case by case first (\ref{geo11}) and then (\ref{geo9}) for the four classes one finds:
\pagebreak
\\
\underline{\it type I}
\be
\begin{gathered}
A~>~0~,\\
0~\leq ~\frac{\sqrt{B^2-4AC}-B}{2A}~\leq ~\sinh ^2\rho ~,
\end{gathered}
\label{geo15}
\ee
\\
\be 
\begin{aligned}
\rho &=\arsh\sqrt{\frac{1}{4A}\left (e^{\pm 2\sqrt A(\tau +\tau _0)}+
(B^2-4AC)e^{\mp 2\sqrt A (\tau +\tau _0)}\right )-\frac{B}{2A}}~,
\\
t&=\pm \arctan\left (\frac{e^{\pm 2\sqrt A(\tau +\tau _0)}+2A-B}{2b\sqrt A}\right )~+~t_0~,\\
f&=\pm \arctan\left (\frac{e^{\pm 2\sqrt A(\tau +\tau _0)}-B}{2\tilde b\sqrt{A}}\right )~+~f_0~,
\end{aligned}
\label{geo19}
\ee
\\[2mm]
\underline{\it type~II}
\be
\begin{gathered}
A~<~0~,~~~~~B^2-4AC~>~0~,~~~~~B~>~0~,\\
0~\leq ~\frac{B-\sqrt{B^2-4AC}}{-2A}~\leq ~\sinh ^2\rho ~\leq ~\frac{B+\sqrt{B^2-4AC}}{-2A}~,
\end{gathered}
\label{geo16}
\ee
\\
\be
\begin{aligned}
\rho &=\arsh\sqrt{\frac{1}{-2A}\big (B\pm\sqrt{B^2-4AC}\sin(2\sqrt{-A}(\tau +\tau _0))
\big )}~,\\
t&=\pm\frac{1}{2}~\arccot\left ( \frac{2\sqrt{-A}~b\cos (2\sqrt{-A}(\tau +\tau _0))}{
\sqrt{B^2-4AC}\pm (B-2A)\sin (2\sqrt{-A}(\tau +\tau _0))}  \right )~+~t_0~,\\
f&=\pm\frac{1}{2}~\arccot\left ( \frac{2\sqrt{-A}~\tilde b\cos (2\sqrt{-A}(\tau +\tau _0))}{
\sqrt{B^2-4AC}\pm B\sin (2\sqrt{-A}(\tau +\tau _0))}  \right )~+~f_0~,
\end{aligned}
\label{geo21}
\ee
\\[2mm]
\underline{\it type III}
\be
\begin{gathered}
A~=~0~,~~~~~B~>~0~,\\
0~\leq ~\frac{-C}{B}~\leq ~\sinh ^2\rho~,
\end{gathered}
\label{geo17}
\ee
\\
\be
\begin{aligned}
\rho&=\arsh\sqrt{B(\tau +\tau _0)^2-\frac{C}{B}}~,\\
t&=\arctan\left (\frac{B(\tau +\tau _0)}{b}\right )~+~t_0~,\\
f&=\arctan\left (\frac{B(\tau +\tau _0)}{\tilde b}\right )~+~f_0~,
\end{aligned}
\label{geo23}
\ee
\\[2mm]
\underline{\it type~IV}
\be
\begin{gathered}
A~<~0~,~~~~~B^2-4AC~=~0~,~~~~B~\geq ~0~,\\
\sinh ^2\rho~=~\frac{B}{-2A}~,
\end{gathered}
\label{geo18}
\ee
\\
\be
\begin{aligned}
\rho&=\arsh\sqrt{\frac{B}{-2A}}~,\\
t&=\sqrt{-A}~\tau ~+~t_0~,\\
f&=\pm\sqrt{-A}~\tau ~+~f_0~.   
\end{aligned}
\label{geo25}
\ee
\\[2mm]

Perhaps it is useful to stress, that in the absence of any movement in
$\Omega _3$, i.e. for $C=0$, the formulas (\ref{geo19}), (\ref{geo21}) and
(\ref{geo23}) for $\rho $ simplify to\\[2mm]
\underline{\it type I with C=0}
\be
\rho ~=~\arsh\left (\sqrt{\frac{\vert B\vert}{A}}~\big\vert\sinh \big (\sqrt{A}(\tau +\tau '_0)\big )\big\vert\right )~,
\label{geo27}
\ee
\underline{\it type II with C=0}
\be
\rho ~=~\arsh\left (\sqrt{\frac{B}{-A}}~\big\vert\sin \big (\sqrt{-A}(\tau +\tau '_0)\big )\big\vert \right )~,
\label{geo28}
\ee
\underline{\it type III with C=0}
\be
\rho ~=~\arsh \left (\sqrt B ~\vert \tau +\tau _0\vert\right )~.
\label{geo29}
\ee

The $\pm$ alternative in (\ref{geo19}) and (\ref{geo21}) has been
absorbed into the shift of the integration constant $\tau_0$ to $\tau'_0$.

The causal properties of the geodesics and their relation to the conformal
boundary (note footnote 5) can be summarized in the following table.\\[3mm]

\begin{tabular}{|p{10mm}|p{40mm}|p{40mm}|p{40mm}|}
\hline
type&causal propert. w.r.t. $AdS_5$&causal propert. w.r.t. $AdS_5\times S^5$&reaches conf. bound. of $AdS_5\times S^5$\\
\hline
\hline
I&space-like&space-like&yes\\
\hline
II&time-like&all&no\\
\hline
III&null&null or space-like&yes\\
\hline
IV&time-like&all&no\\
\hline
\end{tabular}\\[3mm]
 
For later use it is important to stress, that null geodesics in the sense of full $AdS_5\times S^5$ reaching the boundary have to be of {\it type III}. For them no movement
in $S^5$ is allowed while a movement in $S^3$ is possible  
as long as $b^2>\tilde b^2$.
\section{Geodesics in the plane wave}
Here the metric is
\be
ds^2~=~-4dx^+dx^-~-~\vec x^{~2}(dx^+)^2~+~(d\vec x)^2~,
\label{30}
\ee
and yields the geodesic equations
\bea
\ddot x^+&=&0~,\label{geo31}\\
\ddot x^-~+~\frac{1}{2}~\dot x^+\frac{d}{d\tau }\vec x^{~2}&=&0~,\label{geo32}\\
\ddot x^i~+~(\dot x^+)^2x^i&=&0~.\label{geo33}
\eea
(\ref{geo31}) implies linear dependence of $x^+$ on the the affine parameter
$\tau $
\be
x^+~=~\alpha~\tau ~+~x_0^+~.
\label{geo34}
\ee
Obviously now the geodesics fall into two classes, {\it type A} with $\alpha =0$ and {\it type B} with $\alpha\neq 0$.\\[2mm]
\underline{\it type A}\\[2mm]
\be
x^+~=~\mbox{const}~,~~~x^-~=~\beta\tau ~+~x^-_0~,~~~x^i~=~\gamma ^i\tau ~+~x^i_0~.
\label{geo35}
\ee
The scalar product of their tangential vector with itself is given by $(\gamma ^i)^2$.
This implies:

All {\it type A} geodesics are null or space-like. 
Space-like {\it type A} geodesics
reach infinity in the transversal coordinates $\vec x$. {\it Type A} null 
geodesics are given by constant $x^+$ and $x^i$ as well as $x^-$ running 
between
$\pm\infty $.\\[2mm]
\underline{\it type B}\\[2mm]  
Then we have (\ref{geo34}) with $\alpha\neq 0$, and the
integrations of (\ref{geo32}),(\ref{geo33}) yield in addition
\bea
x^i&=&\beta ^i\sin\big (\alpha (\tau +\tau _0^i)\big )~,\label{geo36}\\
x^-&=&\frac{1}{8}\sum _i(\beta ^i)^2\sin\big (2\alpha (\tau +\tau ^i_0)\big )~+~\gamma\tau ~+~x^-_0~.\label{geo37}
\eea
The scalar product of the tangential vector with itself is now equal to $-4\alpha\gamma $,
and we conclude:

All {\it type B} geodesics either stay at $\vec x = 0$ (for $\vec{\beta}=0$) or 
oscillate in the transversal coordinates $x^i$ (for $\vec{\beta}\neq 0$).
All space or time-like {\it type B} geodesics ($\gamma\neq 0$) reach $\pm\infty $
both in $x^+$ and $x^-$. {\it Type B} null geodesics ($\gamma =0$) reach $\pm\infty $ $only$ with respect to $x^+$. Furthermore, they stay at fixed $\vec x$ and $x^-$ ($\vec{\beta}=0$)  or oscillate both in $\vec x$ and
$x^-$ ($\vec{\beta}\neq 0$).

In conclusion null geodesics reaching the conformal boundary of the plane
wave, see (\ref{bound11}), (\ref{bound12}), are necessarily of {\it type A}.
There are no null geodesics reaching the conformal boundary within the
asymptotic regime of limit (i).  \\

Closing this section we comment on a simple discussion of the plane wave 
null geodesics in using the BN coordinates of (\ref{bound5}). In general
null geodesics are invariant under a Weyl transformation. Such a 
transformation only effects the choice of affine parameters along the null 
geodesics. Null geodesics with respect to (\ref{bound5}) without the Weyl
factor are given by great circles in $S^9$ accompanied by a compensating
movement along the time-like direction $\varphi $. If we discuss $S^9$ as
an embedding in $R^{10}$, reaching $\alpha =\frac{\pi}{2}$ 
is equivalent to reaching the $(z_1,z_2)$-plane. There are of course great
circles within this plane. They correspond to null geodesics either winding at
 $\alpha =\frac{\pi}{2}$ in constant distance to the conformal plane wave
boundary around the cylinder in fig.~1 up to $\varphi\rightarrow\pm\infty $ 
or they wind in the orthogonal direction crossing the conformal plane wave
boundary. In the sense of $R\times S^9$ there is nothing special 
with such a crossing. But going back to the metric including the Weyl factor, starting from an inside point, the boundary is reached at infinite affine
parameter. Furthermore, there are of course great circles staying completely
away from the $(z_1,z_2)$-plane (i.e. $\alpha =\frac{\pi}{2}$). They 
correspond to null geodesics generically oscillating in $0<\alpha <\frac{\pi}{2}$ and running up to $\varphi \rightarrow\pm\infty $. Finally, great
circles can also intersect the $(z_1,z_2)$-plane. Then they correspond to
null geodesics oscillating in $\alpha $ and touching $\alpha =\frac{\pi}{2}$.
Obviously some of them reach the conformal boundary line of the plane
wave. According to the above analysis in BMN coordinates they are of
{\it type A}, too. 
\section{Conformal boundaries and geodesics}
As discussed in section 3, only null-geodesics of {\it type III} reach the conformal boundary of $AdS_5\times S^5$. They necessarily stay at fixed $S^5$-position. Translating (\ref{geo23}) into the coordinates of (\ref{intro3})
we get 
\be
\begin{aligned}
x^+&=\frac{1}{2}\left (\arctan\big (\frac{B(\tau +\tau _0)}{b}\big )~
+~t_0~+~\psi\right )~,\\
x^-&=\frac{R^2}{2}\left (\arctan\big (\frac{B(\tau +\tau _0)}{b}\big )~
+~t_0~-~\psi\right )~,\\
r&=R~\arsh\sqrt{B(\tau +\tau _0)^2-\frac{C}{B}}~,\\
f&=\arctan\big (\frac{B(\tau +\tau _0)}{\tilde b}\big )~
+~f_0~,\\
y&=R~\theta ~.
\end{aligned}
\label{bg1}
\ee
Our goal is to find in the $R\rightarrow\infty$ limit a correspondence
to null geodesics of the plane wave. Therefore, our $AdS_5\times S^5$
geodesics have to stay at least partially within the range of finite
$x^+,x^-,r,y$. Taking $R\rightarrow\infty$ at fixed $\tau $ would send all
$x^-$ to infinity. But of course the affine parameter itself is determined
only up to a constant rescaling. Therefore, the best procedure 
is to eliminate the affine parameter completely.

First from (\ref{bg1}) we conclude, that along the full range of a {\it type III} null geodesic, i.e. for
($-\infty<\tau <\infty$), the coordinate $x^+$ runs within an interval of length $\frac{\pi}{2}$:
 $x^+\in \big (\frac{1}{2}(t_0+\psi )-\frac{\pi}{4},\frac{1}{2}(t_0+\psi )+\frac{\pi}{4}\big )$ and $x^-$ runs  within an interval of length $\frac{\pi}{2}R^2$:
$x^-\in \big (R^2(\frac{t_0-\psi }{2}-\frac{\pi}{4}),R^2(\frac{t_0-\psi }{2}+\frac{\pi}{4})\big )$. To ensure that the $x^-$ interval for $R\rightarrow\infty $ stays at least partially within the range of finite values both endpoints of the interval have to have the opposite sign. Thus we have
to restrict $t_0$ and $\psi $ by
\be
-~\frac{\pi}{2}~<~t_0~-\psi ~<~\frac{\pi}{2}~.
\label{bg2}
\ee
In addition one has universally
\be
\vert f(\tau =+\infty )~-~f(\tau =-\infty )\vert ~=~\pi ~.
\label{bg3}
\ee
From (\ref{geo7}) $f$ can be understood as the angle along the great
circle in $S^3$ on which our null geodesics is running. Therefore, 
for {\it type III} geodesics the positions for $\tau =-\infty $
and $\tau =+\infty $ within the $S^3$ are always antipodal to each another.
\footnote{In the limiting case, where the null geodesics goes through $r=0$, $f$ becomes a step function.}

After these preparations we now eliminate the affine parameter and
express $x^+$, $r$ and $f$ in terms of $x^-$
(note that for {\it type III} we have $A=0$ and $B=b^2-\tilde b^2$)
\be
\begin{aligned}
x^+&=\frac{x^-}{R^2}~+~\psi ~,\\
r&=R~\arsh\sqrt{\frac{\tan ^2(\frac{2x^-}{R^2}-t_0+\psi )~+~\frac{\tilde b^2}{b^2}}{1~-~\frac{\tilde b^2}{b^2}}}~,\\
f&=f_0~+~\arctan\left (\frac{b}{\tilde b}\tan\big (\frac{2x^-}{R^2}-t_0+\psi \big )\right )~,\\
y&=R~\theta ~.
\end{aligned} 
\label{bg4}
\ee  
The minimal value for $r$ is 
$$r_{\mbox{\scriptsize min}}~=R~\arsh \left (\frac{\vert\frac{\tilde b}{b}\vert
}{\sqrt{1-\frac{\tilde b^2}{b^2}}}\right )~.$$
Since we insist on finite $r_{\mbox{\scriptsize min}}$ for $R\rightarrow\infty $ we have to rescale ($r_0=\lim _{R\rightarrow\infty}r_{\mbox{\scriptsize min}}$)
\be
\frac{\vert\tilde b\vert }{\vert b\vert }~=~\frac{r_0}{R} ~.
\label{bg5}
\ee
Although we have now realized finite $r_{\mbox{\scriptsize min}}$, the
$x^-$ value where $r_{\mbox{\scriptsize min}}$ is reached stays finite for 
$R\rightarrow\infty$ only
if (\ref{bg2}) is replaced by the stronger rescaling condition
\be
t_0~-~\psi ~=~\frac{a}{R^2}~.
\label{bg6}
\ee

Altogether, to stay at least with part of the {\it type III} null geodesics
within the range of finite BMN coordinates, it is mandatory to perform
the rescalings (\ref{bg5}), (\ref{bg6}) and to keep $y$ fixed. The
remaining parameters replacing $t_0,\psi ,b,\tilde b,f_0,\theta $
are $\psi , a,b,r_0,f_0,y$.

Considering now at fixed $x^-$ the $R\rightarrow\infty$ limit of (\ref{bg4})
one arrives at
\be
\begin{aligned}
x^+&=\psi ~+~O\big (\frac{1}{R^2}\big )~,\\
r&=r_0~+~O\big (\frac{1}{R^2}\big )~,\\
f&=f_0~+~O\big (\frac{1}{R^2}\big )~,\\
y&=\mbox{const}~.
\end{aligned}
\label{bg7}
\ee 
Constant $r,y$ via (\ref{bound6}) give constant $x$. In addition, constant
$f$, i.e. no movement in the $S^3$, and the a priori absence of any movement in $S^5$ leads to constant $\vec x$. This together with the constancy of
$x^+$ implies:

An $AdS_5\times S^5$ null geodesics, reaching the conformal boundary,
for any finite $x^-$-interval at $R\rightarrow\infty $ converges uniformly to a {\it type A} null geodesics of the plane wave.

However, the approach of the $AdS_5\times S^5$ null geodesics to the
conformal boundary of $AdS_5\times S^5$ is realized within the
asymptotic regime of limit (i), see (\ref{bound11}), (\ref{bg1}),  but that of the plane wave null geodesics within the regime of limit (ii), see (\ref{bound12}) and text after (\ref{geo35}). That means even for large $R$, after a
region of convergence, on their way to the boundary they diverge from one another at the very end (in the $x^+,x^-,x$ coordinates under discussion).

In a global setting the situation is most simply illustrated for {\it type III}
null geodesics crossing the origin of the transverse BMN coordinates $\vec x$,
i.e. $r_0=y=0$. We also put $a=0$, the case $a\neq 0$ can be simply recovered
by the replacement $x^-\rightarrow x^--\frac{a}{2}$. Then first
of all $x^-$ runs between $\pm ~\frac{\pi}{4}~R^2$. Furthermore, (\ref{bg4}) implies ($\Theta (z)$ step function)
\be
\begin{aligned}
\frac{r}{R}&=F\big (\frac{x^-}{R^2}\big )~,~~~\mbox{with}~~F(z)~=~\arsh\vert\tan(2z)\vert ~,\\
f&=f_0~\pm~\frac{\pi}{2}~\big (2\Theta (x^-) -1\big )~.
\end{aligned}
\label{bg8}
\ee
The plane wave geodesic is at $x=y=r=0,~~-\infty <x^-<\infty $. It is the uniform limit for $R\rightarrow\infty $ in the region $\vert x^-\vert ~<~R^{1-\epsilon}$. This convergence is due to the different powers of $R$ on the l.h.s. and
in the argument of the function $F$ on the r.h.s. of (\ref{bg8}), see also fig.~3.

The picture in fig.~3 has to be completed by the freedom to choose a point
on $S^3$ to fix the direction in the space of the $\vec x$ coordinates. 
This completely specifies the
{\it type III} null geodesics under discussion. Then the conformal boundary
reaching null geodesics of $AdS\times S^5$ crossing the origin of the
transversal BMN coordinates $\vec x=0$ form a cone with base $S^3$. 
The three parameters
to specify the $S^3$ position together with $\psi $ nicely correspond to the four-dimensionality of the $AdS\times S^5$ boundary. In the $R\rightarrow\infty $
limit this cone degenerates.  
\begin{figure}
\begin{center}
\mbox{\epsfig{file=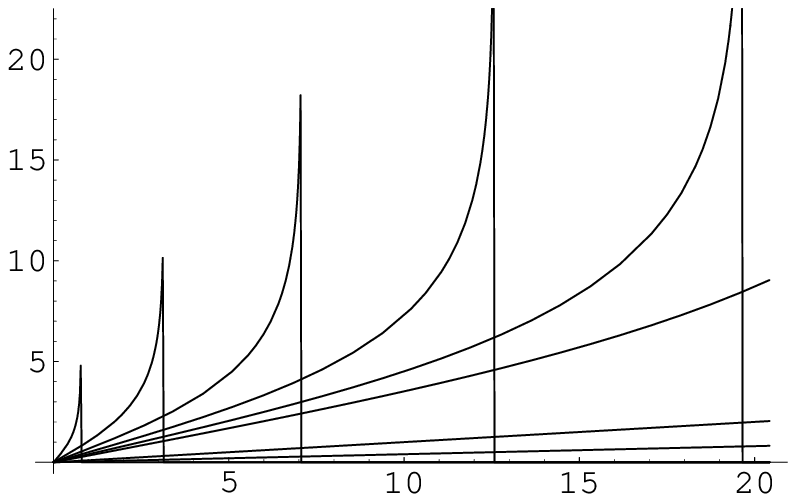, width=100mm}}
\end{center}
\noindent {\bf Fig.~3}\ \ {\it
Approach of boundary reaching $AdS_5\times S^5$ null geodesics to a boundary
reaching null geodesics of the plane wave. The plane wave null geodesics
runs along the horizontal axis up to infinity. The plot shows $r$ versus $x^-$
for $AdS_5\times S^5$ null
geodesics (\ref{bg8}) in the cases $R=1,~2,~3,~4,~5,~6,~20,~50$.}
\end{figure}
\section{Conclusions}
Using BN coordinates the conformal boundary of the plane wave (\ref{intro4})
has been identified as an one dimensional line \cite{bn}. Drawing three
of the BN coordinates ($\varphi ,\alpha ,\beta $), the plane wave is mapped
to a cylinder of infinite length and radius $\frac{\pi}{2}$. To any point
inside this cylinder belongs a $S^7$. These $S^7$ degenerate to a point
on the tube of radius $\frac{\pi}{2}$. The conformal boundary line
spirals around the tube of radius $\frac{\pi}{2}$.

In this paper we have shown that using the same BN coordinates for
$AdS_5\times S^5$ its well known conformal boundary, in the projection
to the three coordinates ($\varphi ,\alpha ,\beta $), appears to be 
located at the same spiraling line as the conformal boundary of
the plane wave. Of course for $AdS_5\times S^5$ on this line the extension 
with respect to the other 7 coordinates is not degenerated to a point.
But we have generated a perhaps useful intuitive picture: The boundary
is always at the same line, taking the limit $R\rightarrow\infty $
the extension in the remaining 7 coordinates shrinks to a point.
\footnote{This then implies also the degeneration of the 3 remaining
dimensions of the conformal boundary of $AdS_5\times S^5$.}

Switching from BN coordinates back to BMN coordinates, it turned out
that, due to the singularity of the coordinate transformation
at the boundary line, the approach to this line is realized within
two different asymptotic regimes of the BMN coordinates, called (i) and (ii) 
in (\ref{bound11}), (\ref{bound12}). Only one of these regimes corresponds to the
conformal boundary of $AdS_5\times S^5$.

Insight into the causal relations between bulk and boundary in the limiting
process can be obtained by analyzing the behaviour of geodesics reaching
the respective boundary. 

We have given a complete classification
for {\it all} geodesics, both for the original full $AdS_5\times S^5$ and
the plane wave. This classification is based on different ranges for
three specific integration constants. Among the $AdS_5\times S^5$ 
null geodesics only those of {\it type III} reach the conformal boundary.
In BMN coordinates this approach is within the asymptotic regime (i).
In contrast, null geodesics of the plane wave reaching the conformal
plane wave boundary approach its boundary within regime (ii).
First of all this obviously implies that for $R\rightarrow\infty $
the convergence of $AdS_5\times S^5$ geodesics to plane wave geodesics
cannot be uniform. Using our explicit formulas for the geodesics
we were able to discuss the issue of convergence in more detail.

The convergence is uniform in the region $\vert x^-\vert <R^{1-\epsilon }$.
Beyond this region, at any fixed $R$, the null geodesics of $AdS_5\times S^5$
and the plane wave diverge while approaching the respective conformal
boundary. Then the plane wave null geodesics runs up to $\vert x^-\vert
\rightarrow\infty $ but stays at finite values for the remaining coordinates.
On the other side, the  $AdS_5\times S^5$ null geodesics asymptotes
to $x^-=\pm\frac{\pi}{4}R^2$ while some of the coordinates different
from $x^+$ and $x^-$ diverge. The fact of converging geodesics within
$\vert x^-\vert <R^{1-\epsilon }$ fits into the naive picture
that in BMN coordinates the $AdS_5\times S^5$ space up to the order
of magnitude of $R$ looks like a plane wave.

The analysis of global properties of the null geodesics of $AdS_5\times S^5$ crossing the origin of the transversal part of the BMN coordinates, given at the end of section 5, can be straightforwardly generalized to geodesics passing this origin at a nonzero distance. Therefore, we conclude that at each point with finite BMN coordinates the null geodesics of $AdS_5\times S^5$ reaching the conformal boundary form a three-dimensional cone. For $R\to\infty$, in the range where the BMN coordinates stay fixed or grow slower than $R$, this cone degenerates to the single plane wave null geodesic crossing the point under consideration and reaching the plane wave conformal boundary. Therefore all points in this range effectively notice a degeneration of the boundary.  
\\[15mm]
{\bf Acknowledgements:}\\[2mm]
This work is supported by DFG (German Science Foundation) within
the ``Schwer\-punkt\-programm Stringtheorie'' and the ``Graduiertenkolleg 271''.
We thank N. Beisert, G. Curio, N. Prezas and M. Salizzoni for useful discussions.


\end{document}